\documentclass[pre,twocolumn,groupedaddress,showpacs]{revtex4}
\usepackage{graphicx}
\begin{document}
\title{Power-law distributions for the areas of the basins of attraction
on a potential energy landscape}
\author{Claire P.~Massen}
\affiliation{Department of Chemistry, University of Cambridge, 
Lensfield Road, Cambridge CB2 1EW, United Kingdom}
\author{Jonathan P.~K.~Doye}
\email{jonathan.doye@chem.ox.ac.uk}
\address{Physical and Theoretical Chemistry Laboratory,
Oxford University, South Parks Road, Oxford OX1 3QZ, United Kingdom}
\date{\today}

\begin{abstract}
Energy landscape approaches have become increasingly popular for analysing
a wide variety of chemical physics phenomena.
Basic to many of these applications has been the inherent structure
mapping, which divides up the potential energy landscape into basins of 
attraction surrounding the minima.
Here, we probe the nature of this division by introducing a method to
compute the basin area distribution and applying it to some archetypal
supercooled liquids.
We find that this probability distribution is a power law over a 
large number of decades with the lower-energy minima having larger
basins of attraction. 
Interestingly, the exponent for this power law is approximately the same
as that for a high-dimensional Apollonian packing, providing further support 
for the suggestion that there is a strong analogy between the way the energy 
landscape is divided into basins, 
and the way that space is packed in self-similar, space-filling hypersphere 
packings, such as the Apollonian packing. 
These results suggest that the basins of attraction provide a fractal-like 
tiling of the energy landscape, and that a scale-free pattern of connections 
between the minima is a general property of energy landscapes.
\end{abstract}

\maketitle 

The potential energy surface, which defines how the potential energy 
depends on the coordinates of all the atoms in a system, has a 
complex, multi-dimensional landscape \cite{Wales03}.
In recent years, there have been intensive efforts to understand the
behaviour of systems, such as proteins, supercooled liquids and clusters,
in terms of features of these `energy landscapes' \cite{Brooks01}.
This research programme has led to important new insights into 
how proteins fold \cite{Dobson98} and the origins of the unusual dynamic
properties of supercooled liquids \cite{Still95,Debenedetti01}.

Many of these applications rely on the `inherent structure' mapping 
introduced by Stillinger and Weber \cite{StillW84a} that is illustrated 
in Fig.\ 1.
It divides the energy landscape into basins of attraction 
surrounding the minima on the energy landscape, where a basin
is defined as the set of points for which following the steepest-descent paths
downhill from those points leads to the same minimum.
The utility of the inherent structure approach is that it allows the 
behaviour of the complete landscape to be conceived in terms of the 
properties of the individual basins \cite{Sciortino05}, which are themselves
tractable to calculate. 
Hence, landscape-based descriptions of a system's
thermodynamics and dynamics can be obtained \cite{Wales03}.

In many of the analyses of energy landscapes, the aim has been to understand 
differences in behaviour, e.g.\ proteins that are good or bad folders 
\cite{Sali94a,Bryngelson95}, 
or supercooled liquids that show `strong' or `fragile' dynamics \cite{Sastry01,Ruocco04},
in terms of differences in the energy landscape. 
However, surprisingly little is known about the 
more universal organizing principles that are common to all such complex 
landscapes. For example, it is known that the number
of minima scales exponentially with system size \cite{Still99}, 
and the distribution of minima as a function of energy is a 
Gaussian \cite{Sciortino99a}, 
but what is the nature of the division of the landscape into basins
and the pattern of connectivities between these basins?
In particular, does the inherent structure mapping lead to an equitable 
division of configuration space into basins, or an inequitable one where
a small minority of the basins occupy a substantial majority of the 
space?

\begin{figure}
\begin{center}
\includegraphics[width=8cm]{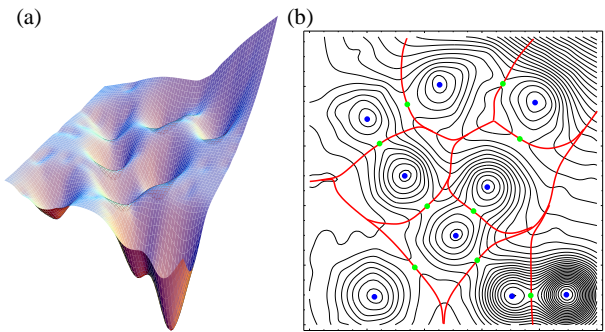}
\end{center}
\caption{\label{fig:IS} (Color online)
(a) A model two-dimensional energy landscape and (b) its
associated contour plot illustrating the inherent structure mapping. 
In (b) the landscape has been divided into basins of attraction 
where the basin boundaries are represented by the thick lines, and the minima and
transition states by points. }
\end{figure}

Clues from recent work on the energy landscapes of small Lennard-Jones clusters 
\cite{Doye02c,Doye05b} perhaps suggest the latter scenario.
The network of minima connected by transition states was found to
be scale-free \cite{Barabasi99}, that is the probability distribution 
for the degree (the number of connections to a node in the network) has
a power-law tail. 
Such a feature has been found for many technological, social and
biochemical networks \cite{Albert02,Newman03a}, however, unlike these other
networks where the scale-free behaviour arises from the dynamics
of network growth \cite{Barabasi99}, these `inherent structure networks' are 
static.  Hence the origin of the scale-free topology remains a puzzle.

\begin{figure}
\begin{center}
\includegraphics[width=6cm]{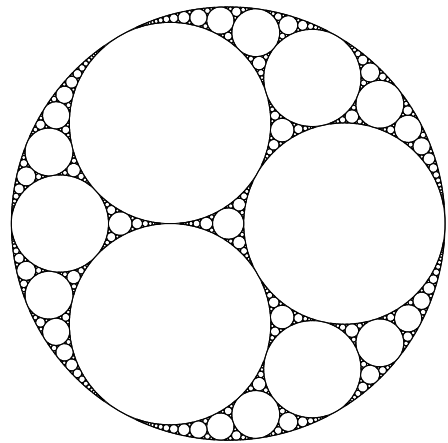}
\end{center}
\caption{\label{fig:Apollo}An Apollonian packing of a circle. Such
a space-filling packing of disks is obtained iteratively, starting
from an initial configuration of three mutually touching disks inside
the circle.
At each iteration disks are added inside each empty curvilinear 
triangle, such that each disk touches all three disks bounding the
triangle. This process is repeated {\it ad infinitum} creating
a space-filling fractal packing of the circle. Its fractal dimension 
is 1.3057 \cite{Manna91}.}
\end{figure}

One potential answer is that the network structure reflects an inequitable
division of configuration space with the low-energy, high-degree minima 
having the largest basins of attraction \cite{Doye98e,Bogdan06}, 
because they are
then more likely to have more transition states along the long basin 
boundaries. Indeed, a recently proposed model spatial, scale-free
network based upon Apollonian packings \cite{Andrade04} has just such a 
correlation between area and degree \cite{Doye05a,Doye05c}.
A two-dimensional example of an Apollonian packing is illustrated in Fig.\ 
\ref{fig:Apollo}. Such a packing is generated by the iterative addition
of ever-smaller disks into the interstices of the packing until the
space is filled. Consequently, the packing is fractal and self-similar \cite{Mandelbrot}.
Importantly, the network of contacts between the disks is scale-free,
and the properties of these `Apollonian networks' are very similar
to the inherent structure networks \cite{Doye05a,Doye05c}.
Thus, there seems to be a potential analogy between the hyperspheres
in a high-dimensional Apollonian packing and the basins
of attraction on an energy landscape. However, the starting point
for this argument was the energy landscapes for clusters with less than 
15 atoms, and it is not clear whether the energy landscapes
for these very small systems are representative.

The fractal nature of the Apollonian packings is evident from the
distribution of the radii of the disks or hyperspheres making up the packing.
This distribution is a power-law at small radii with the exponent related 
to the fractal dimension, $d_f$, where $d-1< d_f < d$ and $d$ is the 
dimension of the space being packed \cite{Boyd73b}. 
This feature potentially allows the 
analogy between the Apollonian packings and the energy landscapes to be
tested further. The distribution of the volumes of the hyperspheres in
an Apollonian packing scales as $V^{-(1+d_f/d)}$, thus leading to the 
prediction that, because of the high dimensionality of typical configuration 
spaces, the distribution of the hyperareas of the basins of attraction 
will follow a power-law with exponent $-2$,
if the basins tile configuration space in an Apollonian-like manner.

To test this prediction we need a method that can obtain this basin area
distribution. To achieve this we study a transformed potential energy 
surface that is commonly used in global optimization 
and is particularly associated with the ``basin-hopping'' 
approach \cite{WalesD97,WalesS99}.
This transformation 
involves assigning the energy at a particular
point in configuration space to that obtained after performing a local
minimization from that point. It transforms the landscape into a set of
steps, where each corresponds to a basin of attraction surrounding a minimum
on the original potential energy surface. 
Thus, for the example landscape in Fig.\ \ref{fig:IS}, the transformed
landscape would consist of nine steps each having the energy of the
corresponding minimum.
Importantly, the probability of sampling 
a step during a simulation on the transformed landscape
is proportional to its hyperarea \cite{Doye98a,Doye98e}. 

Specifically, if we assume that the properties of the basins are 
uniquely characterized by the energy of their minima, 
the probability of being on a step with potential energy in the
range $E\pm dE/2$ in the canonical ensemble obeys
\begin{equation}
p_{\rm min}^{\rm trans}(E,T) dE\propto\Omega_{\rm min}(E) A(E) 
\exp\left(-E/kT\right) dE, 
\end{equation}
where $\Omega_{\rm min}(E)$ is the number of minima with energy
$E\pm dE/2$ and $A(E)$ is the average area of the basins of attraction 
surrounding minima with energy $E$. Hence, we can obtain both $A(E)$ (to
within a multiplicative constant) and the basin
area distribution from $p_{\rm min}^{\rm trans}(E,T)$ distributions obtained
from simulations, if we first know $\Omega_{\rm min}(E)$.
Methods to obtain $\Omega_{\rm min}(E)$ have already been developed 
\cite{Buchner99a,Sciortino99a}, 
and involve inverting similar probability distributions for the original 
potential energy landscape. 
Namely, 
the probability of being in the basin of attraction of a minimum 
with potential energy in the range $E\pm dE/2$ in the canonical 
ensemble is given by
\begin{equation}
p_{\rm min}(E,T) dE\propto\Omega_{\rm min}(E) Z_{\rm vib}(E,T) 
\exp\left(-E/kT\right) dE, 
\end{equation}
where $Z_{\rm vib}(E,T)$ is the vibrational partition function of minima
with energy $E$, which can be calculated, for example, using the harmonic
approximation with frequencies obtained by diagonalization of the 
Hessian matrix of the relevant minima.

Here, we apply this scheme \cite{basinsampling} to three representative 
systems that have been much studied in the supercooled liquids 
community \cite{Sastry98,Dzugutov02b,Barkema98a}, namely 
a binary Lennard-Jones mixture with approximate composition A$_4$B and 
Lennard-Jones parameters as given in Ref.\ \onlinecite{Kob95};
a model one-component glass-forming liquid interacting with the Dzugutov potential 
\cite{Dzugutov91}; 
and amorphous silicon modelled by the Stillinger-Weber potential \cite{StillW85}
with a strengthened 3-body term \cite{Barkema98a}. 
All three systems have 256-atoms and are modelled using periodic boundary 
conditions.

For each system, we ran two sets of simulations.
Firstly, constant-temperature molecular dynamics was performed and 
by regularly minimizing configurations 
generated along the trajectory we obtained $p_{\rm min}(E,T)$ and
hence $\Omega_{\rm min}(E)$ (and a Gaussian fit to it). We chose 
temperatures well above the glass transition where equilibrium sampling
of the liquid is easy to obtain.
Secondly, Metropolis Monte Carlo simulations were performed on the 
transformed landscape, 
from which $p^{\rm trans}_{\rm min}(E,T)$ and hence $A(E)$ was obtained. 
As a check of the sampling, $p^{\rm trans}_{\rm min}(E,T)$ distributions were collected at 
several temperatures. The resulting area distributions were in good agreement,
confirming the reliability of the approach.

\begin{figure}
\begin{center}
\includegraphics[width=8.4cm]{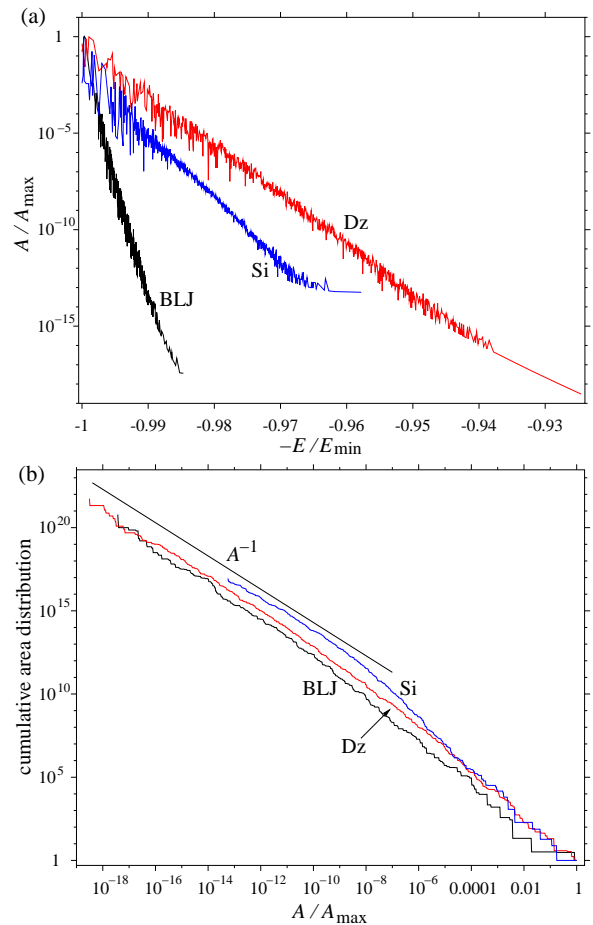}
\end{center}
\caption{\label{fig:pA}  (Color online)
(a) The dependence of the basin area on the energy of the minima, 
and (b) the cumulative basin area distributions 
for the binary Lennard-Jones (BLJ) and Dzugutov (Dz) liquids, and amorphous Si.
To aid comparison between these systems the basin area and the energy are 
measured with respect to the largest basin and the lowest-energy minima that 
have been sampled, respectively.
Additionally, in (b) the power-law predicted by the
analogy to Apollonian packings has been added for comparison \cite{Aminus1}. 
}
\end{figure}

In Fig.\ \ref{fig:pA}(a), $A(E)$ is depicted for these three systems.
In agreement with our expectation that the deeper, more connected minima
should have larger basins of attraction,
the basin areas decrease very 
rapidly with increasing energy in an approximately exponential manner.
The probability
distributions for the hyperareas of the basins of attraction
are illustrated in Fig.\ \ref{fig:pA}(b). It is apparent that the 
distributions follow an approximate power-law over the whole range of areas
sampled, which is between 13 and 18 decades depending on the system.
Closer inspection shows that the distributions begin to curve slightly 
downwards for larger areas, but that for the smaller basins the 
distributions very closely follow the prediction from the Apollonian
analogy for 8 to 9 decades for the binary Lennard-Jones and Dzugutov
systems, and 5 decades for amorphous silicon.

These results have a number of important implications for the 
fundamental properties of energy landscapes, and the inherent structure 
mapping in particular. Firstly, they show that this mapping 
produces a very heterogeneous division of configuration space with
the deeper basins having much larger basins of attraction. 
Secondly, there appears to be a strong analogy between the way that basins of 
attraction tile the energy landscape, and hyperspheres fill space in 
an Apollonian packing. This similarity is exemplified by the power-law
distribution of the basin areas at small $A$, which has exactly the 
expected exponent over a very wide range of basin areas, and 
suggests that the basins divide configuration space in a fractal-like manner. 
Of course, the analogy must break down at some sufficiently small
length scale, because the number of minima on an energy landscape, 
although large, is necessarily finite, whereas the Apollonian packings
require an infinite number of hyperspheres to fill space. However, there
is not yet any sign of this breakdown in the energy range that we have
sampled. We should note that this energy range corresponds to the parts of the
PEL sampled by a liquid, and in future work it would be interesting to try
to use enhanced and biased sampling techniques to sample the basin
area distribution at both higher and lower energies.

Thirdly, if, as for the Apollonian networks, the degree is a power-law 
function of the area \cite{Doye05c} 
(or equivalently the hyperlength of the basin boundary)
the power-law distribution of basin areas is a signature that the underlying
pattern of connections between the minima is scale-free.
Thus, our results suggest that this scale-free topology is not just specific 
to small Lennard-Jones clusters, but a more universal property of energy 
landscapes.
Indeed, there is evidence that the energy landscapes of polypeptides 
also have a scale-free character \cite{Rao04}.
Fourthly, the fractal-like character of the energy landscape provides a static
explanation of the origin of the scale-free topology of inherent structure
networks. However, this is not the end of the matter, since why the basins
provide a fractal-like tiling of the energy landscape is still a puzzle, and one
which we will explore in future work.

\begin{acknowledgments}
The authors are grateful to the Engineering and Physical Sciences Research Council 
(CPM) and the Royal Society (JPKD) for financial support.
\end{acknowledgments}


\begin{thebibliography}{39}
\expandafter\ifx\csname natexlab\endcsname\relax\def\natexlab#1{#1}\fi
\expandafter\ifx\csname bibnamefont\endcsname\relax
  \def\bibnamefont#1{#1}\fi
\expandafter\ifx\csname bibfnamefont\endcsname\relax
  \def\bibfnamefont#1{#1}\fi
\expandafter\ifx\csname citenamefont\endcsname\relax
  \def\citenamefont#1{#1}\fi
\expandafter\ifx\csname url\endcsname\relax
  \def\url#1{\texttt{#1}}\fi
\expandafter\ifx\csname urlprefix\endcsname\relax\def\urlprefix{URL }\fi
\providecommand{\bibinfo}[2]{#2}
\providecommand{\eprint}[2][]{\url{#2}}

\bibitem[{\citenamefont{Wales}(2003)}]{Wales03}
\bibinfo{author}{\bibfnamefont{D.~J.} \bibnamefont{Wales}},
  \emph{\bibinfo{title}{Energy Landscapes}} (\bibinfo{publisher}{Cambridge
  University Press}, \bibinfo{address}{Cambridge}, \bibinfo{year}{2003}).

\bibitem[{\citenamefont{Brooks~III et~al.}(2001)\citenamefont{Brooks~III,
  Onuchic, and Wales}}]{Brooks01}
\bibinfo{author}{\bibfnamefont{C.~L.} \bibnamefont{Brooks~III}},
  \bibinfo{author}{\bibfnamefont{J.~N.} \bibnamefont{Onuchic}},
  \bibnamefont{and} \bibinfo{author}{\bibfnamefont{D.~J.} \bibnamefont{Wales}},
  \bibinfo{journal}{Science} \textbf{\bibinfo{volume}{293}},
  \bibinfo{pages}{612} (\bibinfo{year}{2001}).

\bibitem[{\citenamefont{Dobson et~al.}(1998)\citenamefont{Dobson, \v{S}ali, and
  Karplus}}]{Dobson98}
\bibinfo{author}{\bibfnamefont{C.~M.} \bibnamefont{Dobson}},
  \bibinfo{author}{\bibfnamefont{A.~J.} \bibnamefont{\v{S}ali}},
  \bibnamefont{and} \bibinfo{author}{\bibfnamefont{M.}~\bibnamefont{Karplus}},
  \bibinfo{journal}{Angew. Chem. Int. Edit.} \textbf{\bibinfo{volume}{37}},
  \bibinfo{pages}{868} (\bibinfo{year}{1998}).

\bibitem[{\citenamefont{Stillinger}(1995)}]{Still95}
\bibinfo{author}{\bibfnamefont{F.~H.} \bibnamefont{Stillinger}},
  \bibinfo{journal}{Science} \textbf{\bibinfo{volume}{267}},
  \bibinfo{pages}{1935} (\bibinfo{year}{1995}).

\bibitem[{\citenamefont{Debenedetti and Stillinger}(2001)}]{Debenedetti01}
\bibinfo{author}{\bibfnamefont{P.~G.} \bibnamefont{Debenedetti}}
  \bibnamefont{and} \bibinfo{author}{\bibfnamefont{F.~H.}
  \bibnamefont{Stillinger}}, \bibinfo{journal}{Nature}
  \textbf{\bibinfo{volume}{410}}, \bibinfo{pages}{259} (\bibinfo{year}{2001}).

\bibitem[{\citenamefont{Stillinger and Weber}(1984)}]{StillW84a}
\bibinfo{author}{\bibfnamefont{F.~H.} \bibnamefont{Stillinger}}
  \bibnamefont{and} \bibinfo{author}{\bibfnamefont{T.~A.} \bibnamefont{Weber}},
  \bibinfo{journal}{Science} \textbf{\bibinfo{volume}{225}},
  \bibinfo{pages}{983} (\bibinfo{year}{1984}).

\bibitem[{\citenamefont{Sciortino}(2005)}]{Sciortino05}
\bibinfo{author}{\bibfnamefont{F.}~\bibnamefont{Sciortino}},
  \bibinfo{journal}{J. Stat. Mech.} \bibinfo{pages}{P05015}
  (\bibinfo{year}{2005}).

\bibitem[{\citenamefont{Sali et~al.}(1994)\citenamefont{Sali, Shakhnovich, and
  Karplus}}]{Sali94a}
\bibinfo{author}{\bibfnamefont{A.}~\bibnamefont{Sali}},
  \bibinfo{author}{\bibfnamefont{E.}~\bibnamefont{Shakhnovich}},
  \bibnamefont{and} \bibinfo{author}{\bibfnamefont{M.}~\bibnamefont{Karplus}},
  \bibinfo{journal}{Nature} \textbf{\bibinfo{volume}{369}},
  \bibinfo{pages}{248} (\bibinfo{year}{1994}).

\bibitem[{\citenamefont{Bryngelson et~al.}(1995)\citenamefont{Bryngelson,
  Onuchic, Socci, and Wolynes}}]{Bryngelson95}
\bibinfo{author}{\bibfnamefont{J.~D.} \bibnamefont{Bryngelson}},
  \bibinfo{author}{\bibfnamefont{J.~N.} \bibnamefont{Onuchic}},
  \bibinfo{author}{\bibfnamefont{N.~D.} \bibnamefont{Socci}}, \bibnamefont{and}
  \bibinfo{author}{\bibfnamefont{P.~G.} \bibnamefont{Wolynes}},
  \bibinfo{journal}{Proteins} \textbf{\bibinfo{volume}{21}},
  \bibinfo{pages}{167} (\bibinfo{year}{1995}).

\bibitem[{\citenamefont{Sastry}(2001)}]{Sastry01}
\bibinfo{author}{\bibfnamefont{S.}~\bibnamefont{Sastry}},
  \bibinfo{journal}{Nature} \textbf{\bibinfo{volume}{409}},
  \bibinfo{pages}{164} (\bibinfo{year}{2001}).

\bibitem[{\citenamefont{Ruocco et~al.}(2004)\citenamefont{Ruocco, Sciortino,
  Zamponi, de~Michele, and Scopigno}}]{Ruocco04}
\bibinfo{author}{\bibfnamefont{G.}~\bibnamefont{Ruocco}},
  \bibinfo{author}{\bibfnamefont{F.}~\bibnamefont{Sciortino}},
  \bibinfo{author}{\bibfnamefont{F.}~\bibnamefont{Zamponi}},
  \bibinfo{author}{\bibfnamefont{C.}~\bibnamefont{de~Michele}},
  \bibnamefont{and} \bibinfo{author}{\bibfnamefont{T.}~\bibnamefont{Scopigno}},
  \bibinfo{journal}{J. Chem. Phys.} \textbf{\bibinfo{volume}{120}},
  \bibinfo{pages}{10666} (\bibinfo{year}{2004}).

\bibitem[{\citenamefont{Stillinger}(1999)}]{Still99}
\bibinfo{author}{\bibfnamefont{F.~H.} \bibnamefont{Stillinger}},
  \bibinfo{journal}{Phys. Rev. E} \textbf{\bibinfo{volume}{59}},
  \bibinfo{pages}{48} (\bibinfo{year}{1999}).

\bibitem[{\citenamefont{Sciortino et~al.}(1999)\citenamefont{Sciortino, Kob,
  and Tartaglia}}]{Sciortino99a}
\bibinfo{author}{\bibfnamefont{F.}~\bibnamefont{Sciortino}},
  \bibinfo{author}{\bibfnamefont{W.}~\bibnamefont{Kob}}, \bibnamefont{and}
  \bibinfo{author}{\bibfnamefont{P.}~\bibnamefont{Tartaglia}},
  \bibinfo{journal}{Phys. Rev. Lett.} \textbf{\bibinfo{volume}{83}},
  \bibinfo{pages}{3214} (\bibinfo{year}{1999}).

\bibitem[{\citenamefont{Doye}(2002)}]{Doye02c}
\bibinfo{author}{\bibfnamefont{J.~P.~K.} \bibnamefont{Doye}},
  \bibinfo{journal}{Phys. Rev. Lett.} \textbf{\bibinfo{volume}{88}},
  \bibinfo{pages}{238701} (\bibinfo{year}{2002}).

\bibitem[{\citenamefont{Doye and Massen}(2005{\natexlab{a}})}]{Doye05b}
\bibinfo{author}{\bibfnamefont{J.~P.~K.} \bibnamefont{Doye}} \bibnamefont{and}
  \bibinfo{author}{\bibfnamefont{C.~P.} \bibnamefont{Massen}},
  \bibinfo{journal}{J. Chem. Phys.} \textbf{\bibinfo{volume}{122}},
  \bibinfo{pages}{084105} (\bibinfo{year}{2005}{\natexlab{a}}).

\bibitem[{\citenamefont{Barab\'{a}si and Albert}(1999)}]{Barabasi99}
\bibinfo{author}{\bibfnamefont{A.~L.} \bibnamefont{Barab\'{a}si}}
  \bibnamefont{and} \bibinfo{author}{\bibfnamefont{R.}~\bibnamefont{Albert}},
  \bibinfo{journal}{Science} \textbf{\bibinfo{volume}{286}},
  \bibinfo{pages}{509} (\bibinfo{year}{1999}).

\bibitem[{\citenamefont{Albert and Barab\'{a}si}(2002)}]{Albert02}
\bibinfo{author}{\bibfnamefont{R.}~\bibnamefont{Albert}} \bibnamefont{and}
  \bibinfo{author}{\bibfnamefont{A.~L.} \bibnamefont{Barab\'{a}si}},
  \bibinfo{journal}{Rev. Mod. Phys.} \textbf{\bibinfo{volume}{74}},
  \bibinfo{pages}{47} (\bibinfo{year}{2002}).

\bibitem[{\citenamefont{Newman}(2003)}]{Newman03a}
\bibinfo{author}{\bibfnamefont{M.~E.~J.} \bibnamefont{Newman}},
  \bibinfo{journal}{SIAM Rev.} \textbf{\bibinfo{volume}{45}},
  \bibinfo{pages}{167} (\bibinfo{year}{2003}).

\bibitem[{\citenamefont{Doye et~al.}(1998)\citenamefont{Doye, Wales, and
  Miller}}]{Doye98e}
\bibinfo{author}{\bibfnamefont{J.~P.~K.} \bibnamefont{Doye}},
  \bibinfo{author}{\bibfnamefont{D.~J.} \bibnamefont{Wales}}, \bibnamefont{and}
  \bibinfo{author}{\bibfnamefont{M.~A.} \bibnamefont{Miller}},
  \bibinfo{journal}{J. Chem. Phys.} \textbf{\bibinfo{volume}{109}},
  \bibinfo{pages}{8143} (\bibinfo{year}{1998}).

\bibitem[{\citenamefont{Bogdan et~al.}(2006)\citenamefont{Bogdan, Wales, and
  Calvo}}]{Bogdan06}
\bibinfo{author}{\bibfnamefont{T.~V.} \bibnamefont{Bogdan}},
  \bibinfo{author}{\bibfnamefont{D.~J.} \bibnamefont{Wales}}, \bibnamefont{and}
  \bibinfo{author}{\bibfnamefont{F.}~\bibnamefont{Calvo}}, \bibinfo{journal}{J.
  Chem. Phys.} \textbf{\bibinfo{volume}{124}}, \bibinfo{pages}{044102}
  (\bibinfo{year}{2006}).

\bibitem[{\citenamefont{Andrade et~al.}(2005)\citenamefont{Andrade, Herrmann,
  Andrade, and da~Silva}}]{Andrade04}
\bibinfo{author}{\bibfnamefont{J.~S.} \bibnamefont{Andrade}},
  \bibinfo{author}{\bibfnamefont{H.~J.} \bibnamefont{Herrmann}},
  \bibinfo{author}{\bibfnamefont{R.~F.~S.} \bibnamefont{Andrade}},
  \bibnamefont{and} \bibinfo{author}{\bibfnamefont{L.~R.}
  \bibnamefont{da~Silva}}, \bibinfo{journal}{Phys. Rev. Lett.}
  \textbf{\bibinfo{volume}{94}}, \bibinfo{pages}{018702}
  (\bibinfo{year}{2005}).

\bibitem[{\citenamefont{Doye and Massen}(2005{\natexlab{b}})}]{Doye05a}
\bibinfo{author}{\bibfnamefont{J.~P.~K.} \bibnamefont{Doye}} \bibnamefont{and}
  \bibinfo{author}{\bibfnamefont{C.~P.} \bibnamefont{Massen}},
  \bibinfo{journal}{Phys. Rev. E} \textbf{\bibinfo{volume}{71}},
  \bibinfo{pages}{016128} (\bibinfo{year}{2005}{\natexlab{b}}).

\bibitem[{\citenamefont{Doye and Massen}(2005{\natexlab{c}})}]{Doye05c}
\bibinfo{author}{\bibfnamefont{J.~P.~K.} \bibnamefont{Doye}} \bibnamefont{and}
  \bibinfo{author}{\bibfnamefont{C.~P.} \bibnamefont{Massen}}, in
  \emph{\bibinfo{booktitle}{Complexity, metastability and nonextensivity}},
  edited by \bibinfo{editor}{\bibfnamefont{C.}~\bibnamefont{Beck}},
  \bibinfo{editor}{\bibfnamefont{G.}~\bibnamefont{Benedek}},
  \bibinfo{editor}{\bibfnamefont{A.}~\bibnamefont{Rapisarda}},
  \bibnamefont{and} \bibinfo{editor}{\bibfnamefont{C.}~\bibnamefont{Tsallis}}
  (\bibinfo{publisher}{World Scientific}, \bibinfo{year}{2005}{\natexlab{c}}),
  pp. \bibinfo{pages}{375--384}; cond-mat/0612150.

\bibitem[{\citenamefont{Mandelbrot}(1983)}]{Mandelbrot}
\bibinfo{author}{\bibfnamefont{B.~B.} \bibnamefont{Mandelbrot}},
  \emph{\bibinfo{title}{The Fractal Geometry of Nature}}
  (\bibinfo{publisher}{W. H. Freeman}, \bibinfo{address}{New York},
  \bibinfo{year}{1983}).

\bibitem[{\citenamefont{Manna and Herrmann}(1991)}]{Manna91}
\bibinfo{author}{\bibfnamefont{S.~S.} \bibnamefont{Manna}} \bibnamefont{and}
  \bibinfo{author}{\bibfnamefont{H.~J.} \bibnamefont{Herrmann}},
  \bibinfo{journal}{J. Phys. A} \textbf{\bibinfo{volume}{24}},
  \bibinfo{pages}{L481} (\bibinfo{year}{1991}).

\bibitem[{\citenamefont{Boyd}(1973)}]{Boyd73b}
\bibinfo{author}{\bibfnamefont{D.~W.} \bibnamefont{Boyd}},
  \bibinfo{journal}{Mathematika} \textbf{\bibinfo{volume}{20}},
  \bibinfo{pages}{170} (\bibinfo{year}{1973}).

\bibitem[{\citenamefont{Wales and Doye}(1997)}]{WalesD97}
\bibinfo{author}{\bibfnamefont{D.~J.} \bibnamefont{Wales}} \bibnamefont{and}
  \bibinfo{author}{\bibfnamefont{J.~P.~K.} \bibnamefont{Doye}},
  \bibinfo{journal}{J. Phys. Chem. A} \textbf{\bibinfo{volume}{101}},
  \bibinfo{pages}{5111} (\bibinfo{year}{1997}).

\bibitem[{\citenamefont{Wales and Scheraga}(1999)}]{WalesS99}
\bibinfo{author}{\bibfnamefont{D.~J.} \bibnamefont{Wales}} \bibnamefont{and}
  \bibinfo{author}{\bibfnamefont{H.~A.} \bibnamefont{Scheraga}},
  \bibinfo{journal}{Science} \textbf{\bibinfo{volume}{285}},
  \bibinfo{pages}{1368} (\bibinfo{year}{1999}).

\bibitem[{\citenamefont{Doye and Wales}(1998)}]{Doye98a}
\bibinfo{author}{\bibfnamefont{J.~P.~K.} \bibnamefont{Doye}} \bibnamefont{and}
  \bibinfo{author}{\bibfnamefont{D.~J.} \bibnamefont{Wales}},
  \bibinfo{journal}{Phys. Rev. Lett.} \textbf{\bibinfo{volume}{80}},
  \bibinfo{pages}{1357} (\bibinfo{year}{1998}).

\bibitem[{\citenamefont{B\"{u}chner and Heuer}(1999)}]{Buchner99a}
\bibinfo{author}{\bibfnamefont{S.}~\bibnamefont{B\"{u}chner}} \bibnamefont{and}
  \bibinfo{author}{\bibfnamefont{A.}~\bibnamefont{Heuer}},
  \bibinfo{journal}{Phys. Rev. E} \textbf{\bibinfo{volume}{60}},
  \bibinfo{pages}{6507} (\bibinfo{year}{1999}).

\bibitem[{bas()}]{basinsampling}
\bibinfo{note}{Interestingly, the ``basin-sampling'' approach introduced in
  Ref.\ \onlinecite{Bogdan06} performs the reverse of the current procedure.
  Taking advantage of the enhanced sampling on the transformed landscape
  \cite{Doye98a,Doye98e} and using an approximation for the basin area, basin
  sampling inverts $p_{\rm min}^{\rm trans}(E,T)$ to obtain $\Omega_{\rm
  min}(E)$, and hence the thermodynamic properties of the system. It is
  particularly useful for systems for which ergodicity is hard to achieve on
  the untransformed landscape.}

\bibitem[{\citenamefont{Sastry et~al.}(1998)\citenamefont{Sastry, Debenedetti,
  and Stillinger}}]{Sastry98}
\bibinfo{author}{\bibfnamefont{S.}~\bibnamefont{Sastry}},
  \bibinfo{author}{\bibfnamefont{P.~G.} \bibnamefont{Debenedetti}},
  \bibnamefont{and} \bibinfo{author}{\bibfnamefont{F.~H.}
  \bibnamefont{Stillinger}}, \bibinfo{journal}{Nature}
  \textbf{\bibinfo{volume}{393}}, \bibinfo{pages}{554} (\bibinfo{year}{1998}).

\bibitem[{\citenamefont{Dzugutov et~al.}(2002)\citenamefont{Dzugutov,
  Simdyankin, and Zetterling}}]{Dzugutov02b}
\bibinfo{author}{\bibfnamefont{M.}~\bibnamefont{Dzugutov}},
  \bibinfo{author}{\bibfnamefont{S.~I.} \bibnamefont{Simdyankin}},
  \bibnamefont{and} \bibinfo{author}{\bibfnamefont{F.~H.~M.}
  \bibnamefont{Zetterling}}, \bibinfo{journal}{Phys. Rev. Lett.}
  \textbf{\bibinfo{volume}{89}}, \bibinfo{pages}{195701}
  (\bibinfo{year}{2002}).

\bibitem[{\citenamefont{Barkema and Mousseau}(1998)}]{Barkema98a}
\bibinfo{author}{\bibfnamefont{G.~T.} \bibnamefont{Barkema}} \bibnamefont{and}
  \bibinfo{author}{\bibfnamefont{N.}~\bibnamefont{Mousseau}},
  \bibinfo{journal}{Phys. Rev. Lett.} \textbf{\bibinfo{volume}{81}},
  \bibinfo{pages}{1865} (\bibinfo{year}{1998}).

\bibitem[{\citenamefont{Kob and Andersen}(1995)}]{Kob95}
\bibinfo{author}{\bibfnamefont{W.}~\bibnamefont{Kob}} \bibnamefont{and}
  \bibinfo{author}{\bibfnamefont{H.~C.} \bibnamefont{Andersen}},
  \bibinfo{journal}{Phys. Rev. E} \textbf{\bibinfo{volume}{51}},
  \bibinfo{pages}{4626} (\bibinfo{year}{1995}).

\bibitem[{\citenamefont{Dzugutov and Dahlborg}(1991)}]{Dzugutov91}
\bibinfo{author}{\bibfnamefont{M.}~\bibnamefont{Dzugutov}} \bibnamefont{and}
  \bibinfo{author}{\bibfnamefont{U.}~\bibnamefont{Dahlborg}},
  \bibinfo{journal}{J. Non-Cryst. Solids} \textbf{\bibinfo{volume}{131-133}},
  \bibinfo{pages}{62} (\bibinfo{year}{1991}).

\bibitem[{\citenamefont{Stillinger and Weber}(1985)}]{StillW85}
\bibinfo{author}{\bibfnamefont{F.~H.} \bibnamefont{Stillinger}}
  \bibnamefont{and} \bibinfo{author}{\bibfnamefont{T.~A.} \bibnamefont{Weber}},
  \bibinfo{journal}{Phys. Rev. B} \textbf{\bibinfo{volume}{31}},
  \bibinfo{pages}{5262} (\bibinfo{year}{1985}).

\bibitem[{Ami()}]{Aminus1}
\bibinfo{note}{If the probability distribution scales as $A^{-2}$ then the
  cumulative distribution should scale as $A^{-1}$.}

\bibitem[{\citenamefont{Rao and Caflisch}(2004)}]{Rao04}
\bibinfo{author}{\bibfnamefont{F.}~\bibnamefont{Rao}} \bibnamefont{and}
  \bibinfo{author}{\bibfnamefont{A.}~\bibnamefont{Caflisch}},
  \bibinfo{journal}{J. Mol. Biol.} \textbf{\bibinfo{volume}{342}},
  \bibinfo{pages}{299} (\bibinfo{year}{2004}).

\end{thebibliography}
\end{document}